\documentclass[pre,aps,onecolumn,showpacs,floatfix,superscriptaddress]{revtex4}
\usepackage{graphicx}
\usepackage{dcolumn}
\usepackage{color}
\usepackage{hyperref}
\usepackage{latexsym}
\usepackage{amsmath, amsthm, amssymb}
\usepackage{epsfig}
\usepackage{latexsym}
\usepackage{bm}
\begin{document}

\title{ Strategy switches and  co-action equilibria  in a  minority game }

\author{V. Sasidevan}
\email{sasi@theory.tifr.res.in}
\affiliation{Department of Theoretical Physics, Tata Institute of Fundamental Research, Homi Bhabha Road, Mumbai 400005, India}
\author{Deepak Dhar}
\email{ddhar@theory.tifr.res.in}
\affiliation{Department of Theoretical Physics, Tata Institute of Fundamental Research, Homi Bhabha Road, Mumbai 400005, India}

\date{\today}


\begin{abstract}
We propose an analytically tractable variation of the minority game  in which  rational agents use  probabilistic strategies. In our model, $N$ agents choose between two alternatives repeatedly, and those who are in the minority get a pay-off $1$, others zero. The agents optimize the expectation value of their  discounted future pay-off, the discount parameter being $\lambda$.  We  propose an alternative to the standard Nash equilibrium, called co-action equilibrium,   which gives higher expected pay-off for all agents.  The optimal choice of probabilities of different actions are determined exactly   in terms of   simple  self-consistent equations.   The  optimal strategy is characterized by $N$ real parameters, which  are non-analytic functions of $\lambda$, even for a finite number of agents.  The solution  for $N \leq 7$ is worked out explicitly indicating the structure of the solution for larger $N$. For large enough future time horizon, the optimal strategy switches from random choice to a win-stay 
lose-shift strategy, with the shift probability depending on the current state and $\lambda$.
\end{abstract}

\maketitle

\section{Introduction}
\label{sec1}
There has been a lot of interest in applying  techniques of statistical physics to economics in the past two decades, in particular for a better understanding of  the behaviour of fluctuations in  systems with many   interacting agents, as in a market.  A prototypical model is the El Farol bar problem \cite{elfarol} in which agents optimize their personal pay-offs by guessing what other agents would be doing.  A particular realization of this is the Minority Game (MG) introduced in 1997 by Challet and Zhang \cite{mgpapers1}.   It has  been described  by Arthur as  a  classic `model problem that is simple to describe but offers a wealth of lessons' \cite{foreword}.   In this model, an odd number of agents repeatedly make choices between two alternatives,  and at each step the agents who belong to the minority group are considered as winners. In  MG, the agents cannot communicate with  each other, and base their decision on the common-information which is the history of the winning choices in the last few days.
 
Each agent has a small number of  strategies available with her, and at any time uses her best-performing strategy to decide her immediate future action. The agents are adaptive, and if they find that the strategy they are using is not working well, they will change it. This in turn affects the performance of other agents, who may then change their strategies and so on. Thus, this provides a very simple and instructive  model of learning, adaptation, self-organization and co-evolution in a group of interacting agents. 

Simulation studies of this model showed that  the agents  self-organize into a rather efficient state where there are more winners per day than would be expected if agents made the choice by a random throw of a coin, for a range of the parameters of the model. This  sparked a flurry of interest in the model, and soon after the original paper of Challet and Zhang, a large number of papers appeared, discussing several aspects of the model, or variations. Several good reviews are available in the literature \cite{mg1,mg2}, and there are excellent monographs that describe the known results \cite{mgbook, coolen}. It is one of the few non-trivial models of interacting agents that is also theoretically tractable.

However, one would like to understand  how well a particular strategy for learning and adaptation works, and  compare it with alternate strategies. This strategy to select strategies may be called a meta-strategy. Clearly, the meta-strategy  that gives better pay-off to its user will be considered  better. In this respect, the meta-strategy used in MG does not work so well.  While in some range of parameters, the agents are found to self-organize into a globally efficient state,  in other regions of the parameter space (for  large number of agents),   its overall efficiency is worse than if the agents simply chose the restaurants at random.  This is related to the fact that  in MG,  agents use deterministic strategies, and each agent has only a limited number of deterministic strategies available to her. Also, the rule to select  the strategy to use, in terms of performance scores of strategies, is known to be not very effective. In fact, simulation studies have shown \cite{sornette} that, for some range of parameters, if a small fraction of agents  always select the  strategy   with  the worst performance score,  their average performance,  is better than  of other agents who are using the usual MG rule of choosing the strategy with the best performance score !.

It seems worthwhile, if only to set a point of reference, to determine how well agents in a Minority-like game could do, if they  use some other meta-strategy.  In this paper, we will study a model where the agents use mixed strategies. 
In our formulation, each agent, on each day, selects a probability $p$,  and then  generates a new random number uniformly between $0$ and $1$, and  switches her choice from the previous day if it is $\leq p$. The choice of $p$ depends on the history of the game, and her own history of pay-offs in the last $m$ days, and constitutes the strategy of the agent [The deterministic strategies are  special cases when $p$ is $0$ or $1$]. We discuss how the optimal value of this parameter depends on the history of the game.

For this purpose, we propose a new solution concept, as an alternative to the usual notion of Nash equilibrium \cite{harsanyi}.  We show that the Nash equilibrium  states are  not very satisfactory for our model, giving rise to `trapping states' (discussed below),  and our proposed alternative, to be  called  co-action equilibrium, avoids this problem.  To distinguish it from the original Minority game, we will refer to the new game as the Co-action Minority game (CAMG), and refer to the original MG as Challet-Zhang Minority Game (CZMG).

In CZMG, on any day, each agent selects one strategy from a small set of deterministic strategies given to her at the beginning of the game.  We make the basket of strategies given to the agents much bigger,  and  make all strategies, within a specified  infinite class,  available to all agents.  The use of stochastic, instead of the deterministic, strategies makes the CAMG   more efficient than CZMG. Also, the absence of  quenched disorder - in the form of assigning strategies to agents in the beginning of the game - in our model  makes it much more tractable. One can determine the behaviour of many quantities of interest in more detail, using only elementary algebra. The theoretical analysis of CZMG requires more sophisticated  mathematical techniques such as functional integrals, and taking special limits of  large number of agents, large backward time horizon, and large times (explained later in the paper).

We find that the  optimal strategies of agents can be determined by a mean-field theory like  self-consistency requirement. For the $N$-agents case, we get coupled algebraic   equations in $N$ variables \cite{lnm2003}. The simplicity of our analysis  makes this model an interesting and instructive, analytically tractable model of interacting agents.  Interestingly,  this also  provides us with a non-trivial  example of a non-equilibrium steady state  which shows a non-analytic dependence on a control parameter {\it even for finite number of agents}.

The plan of the paper is as follows: In Sec. \ref{sec2}, we recapitulate the main features of the CZMG, and what is known about its behaviour. In Sec. \ref{sec3},  we introduce  the CAMG game. In Sec. \ref{sec4}, we show that the model has  Nash equilibrium states that are trapping states, where all agents stay with the same choice next day, and the system gets into a frozen state. In Sec. \ref{sec5}, we introduce the solution concept of co-action equilibrium to avoid these trapping states. Sec. \ref{sec6} develops the general theoretical framework of Markov chains to calculate the expected pay-off functions of agents in CAMG, which is used to  determine the optimal strategies by agents. In Sec \ref{sec7}, we work out explicitly, the optimal strategies when the number of agents $N=3, 5$ and $7$, and discuss what one may expect for larger $N$.  In Sec. \ref{sec8}, we discuss the case of large $N$, and study the first transition from random state to one where some of the agents  choose not to jump.  Sec. \ref{sec9} contains a summary of our results, and some concluding remarks.

\section{The  Challet-Zhang Minority Game}  
\label{sec2} 
In CZMG \cite{mgpapers1}, each of the $N$ agents, with $N$ odd, has to choose between two alternatives, say two restaurants A and B, on each day and those in the  restaurant with fewer people get a pay-off $1$, and others $0$. The agents cannot communicate with each other, and  make their choice based only on the information of which was the minority restaurant for each of the last $m$ days. A strategy gives which one of the two choices (A or B) is preferred, for each of the $2^m$ possible histories of the game.  The total number of possible strategies is $2^{{2^m}}$. Each agent has a small fixed number $k$ of strategies randomly picked out of all possible strategies at the beginning of the game. For each of the strategies assigned to an agent, she keeps a performance score which tells how often in the past the strategy correctly predicted the winning choice. On each day, she  decides which restaurant to go to, using the strategy that performed best in the recent past.

We write $N= 2M +1$. Clearly, on any day, the number of people that are happy ( i.e. having a positive pay-off)  is $\leq M$. 
The amount by which the average number of happy people per day differs from the maximum possible value $M$ is a measure of the social inefficiency of the system. For a system of agents in a steady state ${\mathbb S}$,  we will characterize the inefficiency of the system  in terms of  a parameter $\eta$, called the inefficiency parameter, defined as,  
\begin{equation}
\eta_{{\mathbb S}}  =\frac{ W_{max} - \langle W \rangle_{{\mathbb S}} }{W_{max} - W_{rand} },
\label{gain}
\end{equation}
where $W_{max} = M/N$ is the maximum possible pay-off per agent, $\langle W \rangle_{{\mathbb S}} $ is the average pay-off per agent in the steady state ${\mathbb S}$, and $W_{rand}$ is the average pay-off per agent when agents select randomly between A and B.

\begin{figure}
 \includegraphics[scale=.5]{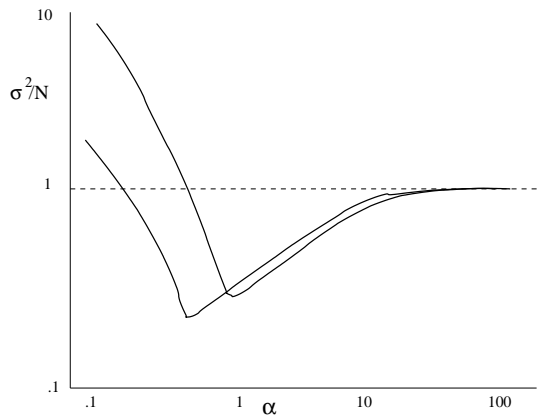}
 \caption{Schematic representation of the variation of the normalized fluctuation in the attendance difference between the two restaurants  $\sigma^2/N$ with the parameter $\alpha = 2^m/N$ for two different values of $k$ - the number of strategies with each agent. The curve with lower minimum corresponds to $k=2$ and the other curve corresponds to $k = 3$. The dashed horizontal line shows the value of $\sigma^2/N$ when agents choose randomly between the two restaurants.} 
 \label{fig1}
\end{figure}

The general  qualitative behaviour of MG is quite well understood from simulations. Fig. \ref{fig1} shows the schematic behaviour of the system as seen in simulations. The theoretical analysis is rather complicated, and involves several limits:  large  $N$ and   $m $, with $ 2^m/N = \alpha $ held fixed.  Also, one has to rescale time with $N$, and the exact theoretical results are possible for fixed $\tau =t/N $.  The asymptotic behaviour in the steady state can be determined exactly only in these limits,  and only for $\alpha$ greater than a critical value $\alpha_c  $, using concepts and formalism developed originally for the spin-glass problem. For a more detailed discussion,  see \cite{mgbook, coolen}.

\section{The Co-action minority game }
\label{sec3} 
We would like to construct a game which preserves the basic simplicity of the Challet-Zhang minority game,  but changes it in several important ways, to make it more tractable. We will keep the allowed actions, and pay-off function the same, but consider different strategies used by agents.   We discuss these changes one by one.  
  
\subsection{Stochastic versus deterministic strategies}

It is well-known that in  repeated games where agents have to make their choices simultaneously, probabilistic strategies are much more effective than deterministic ones. In fact,  Arthur, in the forward of \cite{mgbook}, recalls that when he  introduced the El-Farol Bar problem at a conference, the session chair Paul Krugman had objected that the problem has a simple efficient strategy,  where each agent uses a mixed strategy, and  decides between the two options by tossing a coin.   Of course, in some minority-like games,  like managing  hedge funds, the agents do not have the option of using probabilistic strategies. In the CAMG, we allow the agents to use mixed strategies.   We will show that this results in a different emergent behaviour of the system.

In the CZMG, the agents are assigned a small number of strategies at random, and different agents have different basket of strategies.  In CAMG, we allow each agent to choose his shift probability to be any value $p$, with the only constraint being $0 \leq p \leq 1$. Since the choice of $p$ constitutes the strategy of an agent, each agent is allowed to choose from an infinite set of strategies. Also, the same set of strategies is available to all the agents. Thus we do not have any quenched disorder in the model, and this simplifies the analytical treatment of the model considerably.

Mixed strategies in the context of minority game  have been discussed before.  An example is the thermal minority game \cite{thermal}. A somewhat similar  model to ours, involving a minority game, with probabilistic strategies was studied earlier by Reents et al \cite{reents}. However, there is an important difference between these earlier studies, and ours.  In the earlier studies, the probabilities of different actions was thought to be due to a kind of noise
in the system, not under the control of the agents. The probability of `non-optimal' choice is externally prescribed in the beginning. In our model, the agent is free to choose the value of $p$,  and chooses it to maximize her pay-off. Also, the agent's choice can vary from day to day.

\subsection{Rational versus adaptive agents} 

Another important difference from the CZMG is that we will treat our agents as intelligent and rational who also expect  other agents to act rationally.  The agents  are  selfish, and choose their actions  only to maximize  their personal expected gain. This is an important difference from CZMG, where one of the motivations for introducing the game was to model the behaviour of agents with only bounded rationality who resort to inductive reasoning.  In CZMG, the agents follow rather simple rules to decide when to switch their strategies, based on the performance scores of strategies. One may imagine that these agents are unthinking machines, following some pre-programmed instructions. 

We will also assume that all agents are equally intelligent. Thus, there is no built-in heterogeneity in the behaviour of agents. When the agents follow deterministic strategies as in CZMG, they are forced to differentiate amongst themselves in the strategies they use, as many  agents following the same strategy is clearly not good.  When agents use mixed strategies, the strategies need not be  different, as the differentiation is naturally  provided by the random number generators used by the agents. In fact, this differentiation is rather efficient.  We will see that even without any assumed heterogeneity of agents, the system shows non-trivial emergent behaviour,
and reaches a more efficient state quicker.

We need not discuss here the question whether  full rationality  or  bounded rationality  can describe the behaviour of real-life agents. Clearly, there would be situations where one or the other is a better model.  We only note that  the general  probabilistic  \textquoteleft win-stay lose-shift\textquoteright\; strategy has been seen to be used in many real-world learning situations \cite{nowak}, and this  is also  the strategy  that is found to be optimal by rational agents in CAMG.  

\subsection{Agents' optimization aim}

The next issue is deciding the pay-off function that is optimized by the
rational  agents. Clearly, maximizing the probability of winning next day is a possible goal. But  agents in repeated games need not be concerned only about their immediate payoffs.  

We note that if all agents had the same optimization goal of  minimizing the system inefficiency, there is no competition between them, and we get a  trivial game. There is a simple strategy which will give the best possible result of long-term average of the  inefficiency parameter being zero.  In this strategy, each day, agents  choose some shift probability strictly between $0$ and $1$,  until one finds $M$ people in one restaurant, and $M+1$ in the other. Once this state is reached, each agent goes to the same restaurant  on all subsequent days.  In this simple strategy, the long-time average gain per agent per day is  $M/(2M+1)$, as each agent is equally likely to end up being in the winning set.  

Clearly, this optimization goal makes the game trivial.  A more reasonable goal for selfish agents  would be  to try to optimize their personal long-term average pay-off. Selfish agents in the majority restaurant would not be interested in pursuing the  strategy outlined above. But the game with selfish agents who aim to maximize personal long-term average pay-off is  also easily solved.  If we allow agents in the state with $(M,M+1)$ decomposition to shift with a small probability $\epsilon$, then in the long-time steady state, the system jumps between different decompositions of the type $(M,M+1)$. When $\epsilon $ is small, in the steady state, the state $(M,M+1)$ still occurs with a large weight,  and   this weight will tend to $1$ as $\epsilon $ tends to zero. However, for any non-zero value of $\epsilon$,  for times $T \gg 1/\epsilon$, the fraction of time spent in the minority by each agent  will be nearly the same. Thus, for any specified history, this strategy still gives expected future long-term 
pay-off that tends to the  highest possible, as  $\epsilon$ tends to zero.  

The  problem with the the above game is that, for small $\epsilon$, an agent in the unhappy situation in the $(M,M+1)$ breakup may have to wait very long before her pay-off changes.  This suggests that a reasonable model of the agent behaviour would be that he/she does not want to wait for too long for the next winning day.  We therefore consider agents who have a finite future time-horizon.  Clearly, we can think of agents who try to maximize their net pay-off over the next $H$ days.  

It is more convenient to introduce a real parameter $\lambda$, lying between $0$ and $1$, and  assume that  any agent X only wants to optimize her  weighted expected future pay-off, 
\begin{equation}
{\rm ExpPayoff}(X) = \sum_{\tau=0}^{\infty} [( 1 - \lambda) \lambda^{\tau}] \langle W_X(\tau+1)\rangle,
\end{equation}
where $\langle W_X(\tau)\rangle$ is the expected pay-off of the agent X on the $\tau$-th day ahead, and $\lambda$ is a parameter $0 \leq \lambda < 1$. It is called as the discount parameter in the literature and is easier to deal with than the discrete parameter $H$. 
The factor $(1- \lambda)$ has been introduced in the definition of ${\rm ExpPayoff} (X)$ so that the maximum possible value of the  expected payoff is $1$.

 The advantage of using the real parameter $\lambda$, instead of the discrete parameter $H$, is that we can study  changes in the steady state of the system as we change the parameter $\lambda$ continuously.  We will find that there are  strategy switches in the optimal strategies of agents as $\lambda$ is varied, which leads to discontinuous changes in several properties of the steady state. These are of interest, as they  are analogous to dynamical phase transitions in steady states in non-equilibrium statistical mechanics. 

Consistent with our assumption of homogeneity of agents, we assume that all agents use the  same value of  $\lambda$.

\subsection{Common information}

We assume that all agents know the total number of players $N$, and they also know that all players use the stochastic  strategy of selecting a jump probability based on previous days outcome, and use the same value of discount parameter $\lambda$. 
In the CZMG, on each day, which restaurant was the minority is announced publicly, and this information is  available to all the agents. 
In CAMG, we assume that the common information is more detailed: each agent knows the time-series $\{n(t)\}$ of how many people went to A on different days in the past. We note that in the El Farol bar problem that led to MG,  has the same information as in our variation.  
On any day, the attendance in restaurant A can take $(N+1)$ possible values. Then history of $m$ previous days can take $(N+1)^m$ values, and each strategy is specified by $(N+1)^m$ real numbers.

We restrict our discussion here to the simplest case, where  $m =1 $  for all agents. Then, an agent's strategy is solely determined by the number of people who were in the restaurant she went to the previous day. Since this number cannot be zero, the number of possible histories here  is $N$, not $(N+1)$. 

\section{The problem of trapping states}
\label{sec4} 

This model was first defined in \cite{dhar}. In that paper, we tried to determine the optimal choice of the shift probabilities using the standard ideas of Nash equilibrium.  However, we realized that in this problem, there are special states  such that  none of the agents in that state would prefer to shift to a different restaurant the next day, following a Nash-like analysis. This frozen steady state may be called the trapping state.   

The existence of such a trapping state is paradoxical, as rational agents in the majority restaurant have no reason to pursue a strategy that makes them stay in a losing position for ever. The resolution of this  paradox  requires  a new solution concept, that we discuss now.

The most commonly used notion in deciding optimal strategies in $N$-person games is that of Nash equilibrium: A state of the system in which agent $i$ uses a strategy ${\cal S}_i$ is a Nash equilibrium, if for all $i$, ${\cal S}_i$  is the best response of $i$, assuming that all agents $j \neq i$ use the strategy ${\cal S}_j$.  There may be more than one Nash equilibria in  a given game, and they need not be very efficient. For CAMG also, the Nash equilibrium is not very satisfactory: it gives rise to a trapping state.

Consider, for simplicity, the case $\lambda=0$, where agents optimize only next day's pay-off.
Now, during the evolution of the game, at some time or the other, the system will reach a state with $M$ agents in  one restaurant (assume A), and $M+1$ agents in the other restaurant B.  What is the best strategy of these agents who want to maximize their expected pay-off for the next day? 

We imagine that each agent hires a consultant to advise them. To an agent in A (we will call her Alice),  the advise would be to stay put, if the probability that no person switches from the  restaurant $B$ is greater than 1/2.  If the agents in the  restaurant  $B$ switch with probability $p_{M+1}$, the probability that no one switches is $(1-p_{M+1})^{M+1}$.  In this case, the expected pay-off of Alice would be $(1-p_{M+1})^{M+1}$. So long as this $p_{M+1}$ is small enough that this pay-off is $> 1/2$, Alice's best strategy  would be to choose $p_M =0$.

If an agent in the  restaurant  $B$ (let us call him  Bob)  expects that  agents in $A$ would not switch, what is his best response? The consultant argues that if  Bob  switches, he would be in the majority, and his pay-off would be zero. Hence his best strategy is to set his switching probability  $p_{M+1}$ to zero. Then, there is some possibility that he will be in the winning set the next day, if some other agent from $B$ shifts. In fact, with agents in A staying put ($p_M= 0$), the probability that he wins is proportional to his stay-put probability, and is maximized for $p_{M+1}=0$.

This value $p_{M+1} = p_M = 0$, is  then a self-consistent choice corresponding to the the fact that the choice $p_M=0$ is an individual agent's best response to opposite restaurant's people choosing $p_{M+1} =0$, and vice versa. It is a Nash equilibrium.

This advice is given to all agents in restaurant $B$, and then no one shifts, and the situation next day is the same as before. Thus, the system gets trapped into a  state where all agents stick to their previous day's choice.  In this state, the total number of happy people is the best possible,  and the state has the best possible global efficiency. However, this situation  is very unsatisfactory for {\it the majority of agents} (they  are on the losing side for all future days).  Setting  $p_{M+1}$ equal to zero by agents in B, is clearly not   an optimal choice.  

Let us denote the state of an agent who is in a restaurant with total of $i$ people in it as ${\cal C}_i$.  In the Nash equilibrium  concept, an agent in the state $C_{M+1}$, who believes that agents in the opposite restaurant would be setting their switch probability  $p_M=0$, is advised that his best response is to set $p_{M+1}=0$. If other agents are using $p_{M+1}=0$, one cannot do better by moving. If other agents in the restaurant switch with probability $p_{M+1} \neq 0$, this is the \textquoteleft optimal\textquoteright  ~solution. {\it This does not take into account the fact that if all agents follow this advice, their expected future gain is zero}, which is clearly unsatisfactory: No  other advice could  do worse!

In our previous paper \cite{dhar}, we realized this problem, but could see no way out within the Nash solution concept. We adopted an {\it ad hoc} solution, where the agents in state ${\cal C}_{M+1}$ were required by external fiat to shift with a non-zero probability $\epsilon$.   The alternate solution concept of co-action equilibrium provides a natural and rational solution to this problem. This is explained in the next section.  Note that the presence of even a small number of agents who always choose randomly would   keep the system away from  the trapping state \cite{biswas}.

\section{ An alternate Solution concept: Co-action equilibrium}
\label{sec5} 

The problem with the consultant's reasoning lies in the Nash-analysis assumption of optimizing over strategies of one agent, {\it assuming that other agents would do as before}.  Let the marked agent be denoted by $X$. All agents in the same retaurant, who are not $X$ denoted by $X'$.  Then, the agent $X$ determines his jump probability $p_X$ to optimize his expected payoff ${\rm ExpPayoff} (X) = (1 - p(X)) ( 1 -\prod_{X'} ( 1 - p(X'))$. In this case,  by varying with respect to $p(X)$, keeping all $p(X')$ constant, the payoff is clearly maximized at $p(X)=0$.

 In the alternate co-action equilibrium concept proposed here, an agent in   state $C_i$ realizes that she can choose her switching probability $p_i$, but all the other fully rational $(i-1)$ agents in the same restaurant, with the same information available,  would argue similarly, {\it and choose the same value of $p_i$}. Determining the optimal value of $p_i$ that maximizes the pay-off of agents in state $C_i$ does not need communication between the agents.

If $p_M=0$, then the expected pay-off $W_{M+1}$ of an agent in restaurant B is clearly given by the probability that he does not shift, but at least one of the other agents in his restaurant does. This is easily seen to be  $ q_{M+1} ( 1 - q_{M+1}^M)$, where  $q_{M+1}= 1 -p_{M+1}$.
This is zero for $q_{M+1}= 0$ or $1$, and becomes maximum when $q_{M+1}= (M+1)^{-1/M}$.
In particular, $q_{M+1}$ equal to $1$ is {\em no longer the optimal response}.  

One may argue that this solution concept is not so different from the usual Nash equilibrium, if one thinks of this as a  two-person game each day, where the two persons are the majority and the minority groups, and they select the optimal values of their strategy  parameters $p_i$ and $p_{N -i}$.  The important point is that these groupings are temporary, and  change with time.
For non-zero $\lambda$, one cannot think of this game as a series of two-person games.

  In our model, the  complete symmetry between the agents, and the assumption of  their being fully  rational, ensures that they will reach the co-action equilibrium. 

Note that  an agent in B may wants to \textquoteleft cheat\textquoteright\, by deciding not to shift,  assuming that other agents would shift with a nonzero probability. But this is equivalent to setting his strategy parameter $p=0$. Our assumption of rationality then implies that all other agents, in the same situation, would  argue in the same way, and do the same.



\section{Determining the optimal mixed strategy}
\label{sec6} 
 For a given $N$, a person's full strategy $\mathbb{P}$ is defined by the set of $N$ numbers $\mathbb{P} \equiv  \{p_1,p_2,....p_N\}$. 
In CAMG, all rational agents would end up selecting the same optimal values of strategy parameters $\{p_1^*, p_2^*, \ldots\}$. It would have been  very inefficient for all agents to use the same strategy, if they were using deterministic rules. This is not so in CAMG.
We now discuss the equilibrium choice $\{p_1^*,p_2^*,\ldots p_N^*\}$. The co-action equilibrium condition that $p_i^*$ is chosen to maximize the expected pay-off of agent in state $C_i$,  implies $N$ conditions  on the $N$ parameters $\{p_i^*\}$. There can be more than one self-consistent solution to the equations, and each solution corresponds to a possible steady state.

Clearly, as all agents in the restaurant with $i$ agents switch independently with probability $p_i$, the system undergoes a Markovian evolution, described by a master equation. As each agent can be in one of   the two states, the state space of the Markov chain is $2^N$ dimensional.  However, we use the symmetry under permutation of agents to reduce the Markov transition matrix to $N \times N $ dimensional.
Let $|Prob(t)\rangle$ be an $N$-dimensional vector, whose $j$-th element is $Prob_j(t)$, the probability that a  marked agent $X$ finds herself in  the state  $C_j$ on the $t$-th day.  On the next day, each agent will switch according to the probabilities given by ${\mathbb P}$, and  we get
\begin{equation}
 |Prob(t+1)\rangle = \mathbb{T} |Prob(t)\rangle,
\end{equation}
where $\mathbb{T}$ is the $N\times N$ Markov transition matrix.

Explicit matrix elements are easy to write down. For example, ${\mathbb T}_{11}$ is the conditional probability that  the marked agent will be in  state $C_1$ on  the next day, given that she is in $C_1$ today. This is the sum of two terms: one corresponding to everybody staying
with the current choice  [the probability of this is  $ (1- p_1) (1 -p_{N-1})^{N-1}$], and another corresponding to all agents switching their choice [ the probability is $ p_1 p_{N-1}^{N-1}$].

The total expected pay-off of $X$, given that she is in the state  $C_j$ at time $t=0$ is easily seen to be
\begin{align}
W_j &= (1-\lambda) \left\langle L\left|\dfrac{\mathbb{T}}{1-\lambda \mathbb{T}}\right|j\right\rangle,
\end{align}
where $|j\rangle$ is the column vector with only the $j$-th element $1$, and rest zero; and $\langle L|$ is the left-vector $\langle 1,1,1,1,..0,0,0..|$, with first $M = (N-1)/2$ elements $1$ and rest zero.

In fact, we can use the permutation symmetry between the agents to block-diagonalize the matrix ${\mathbb T}$ into a two blocks of size $(M+1)$ and $M$. This is achieved by a change of basis, from vectors $|i\rangle$ and $|N-i\rangle$ to  the basis vectors $|s_i\rangle$ and $|a_i\rangle$, where 
\begin{eqnarray}
|s_i\rangle = |i\rangle + |N-i\rangle \nonumber, \\
|a_i\rangle = (N-i)|i\rangle - i |N-i\rangle. 
\label{basis}
\end{eqnarray}
This choice is suggested by the fact  that in the steady state
\begin{equation} 
 Prob(C_i)/i = Prob(C_{N-i})/(N-i).
\end{equation}

It is easily verified that  using the basis  vectors $|s_i\rangle$ and $|a_i\rangle$, the matrix ${\mathbb T}$ is block-diagonalized.

One simple choice is that $p_i^* = 1/2$ for all $i$, which is the random choice strategy, where each agent just picks a restaurant totally randomly each day, independent of history. We will denote this strategy by ${\mathbb P}_{rand}$. In the corresponding steady state, it is easy to see that $W_j$ is independent of $j$, and is given by
\begin{equation}
W_j = W_{rand} = 1/2  - \binom{N-1}{M}  2^{-N}, {\rm ~~for ~~all} ~j.
\end{equation}

By the symmetry of the problem, it is clear that $p_N^*=1/2$ for all $\lambda$. Now consider the strategy $\{p_i^*\} = \{p_1^*,1/2,1/2,1/2...\}$.  If X is in  the state $C_1$, and next day all other agents would switch with probability $1/2$,  it does not matter if X switches or not:  payoffs  $W_1$ and $W_{N-1}$ are independent of $p_1^*$.  Hence $p_1^*$ can be chosen to be of any value. It is easy to see that  the strategy  ${\mathbb P}'_{rand}$ in which $p_{1}^* =0$ and $p_{N-1}^* < 1/2$, chosen to maximize $W_{N-1}$, is better for all agents and is stable, and hence is always preferred over ${\mathbb P}_{rand}$.

\section{Exact solution for small \texorpdfstring{$N$}{}}
\label{sec7} 
\subsection{{\bf \texorpdfstring{$N=3$}{}}}

We consider first the  simplest case $N=3$. Since $p_1^*=0$, $p_3^*=1/2$, the only free parameter is $p_2^*$.  The value of $p_2^*$ is decided by the agents in state $C_2$, and they do it by maximizing $W_2$. 

In this case, the transfer matrix is easily seen to be
\begin{equation}
\mathbb{T}=\left[
\begin{array}{ccc}
 q_{2}^2 & p_2 q_2 & 1/4 \\
 2 p_2 q_2  & q_2 & 1/2 \\
 p_{2}^2 & p_{2}^2 & 1/4
\end{array}
\right],
\label{matrix}
\end{equation}
where $q_2 = 1 - p_2$. The pay-off $W_2$ is given by
\begin{align}
\nonumber W_2 &= (1-\lambda) \left[1\; 0\; 0\right]\dfrac{\mathbb{T}}{\left(1-\lambda \mathbb{T}\right)}\left[\begin{array}{c} 0 \\ 1 \\ 0 \end{array}\right], \\
    &= \dfrac{4 p_2 q_2 - \lambda p_2( q_2 -p_2)}{(1 - \lambda q_2 (q_2 - 
       p_2)) (4 + \lambda(4 p_2^2 - 1))}.
       \label{n=3}
\end{align}
The eigenvalues of the transfer matrix  ${\mathbb T}$   are easily seen to be  $\left(1,\;\dfrac{1}{4} \left(1 - 4 p_{2}^2\right),\;q_{2}\left(q_2-p_2\right)\right)$. The  eigen vectors are easily written down. The average gain in the steady state $W_{avg}$ is seen to be 
\begin{equation}
W_{avg} = \dfrac{1}{3 + 4 p_{2}^2}.
\end{equation} 

From Eq. \ref{n=3}, the value of $p_2$ that maximizes $W_2$ is easily seen to be root of the  following cubic equation in $\lambda$.
\begin{align*}
&16 - 32 p_2^* - (24 - 56 p_2^* + 32 p_2^{*2})\lambda +(9-28 p_2^* +40p_2^{*2} \\
&-96p_2^{*3}+144p_2^{*4}-64p_2^{*5})\lambda^2 - (1-4p_2^*+8p_2^{*2}-24p_2^{*3} \\
&+48p_2^{*4}-32p_2^{*5})\lambda^3 = 0.
\end{align*}

\begin{figure*}
\centering
 \includegraphics[scale = .68]{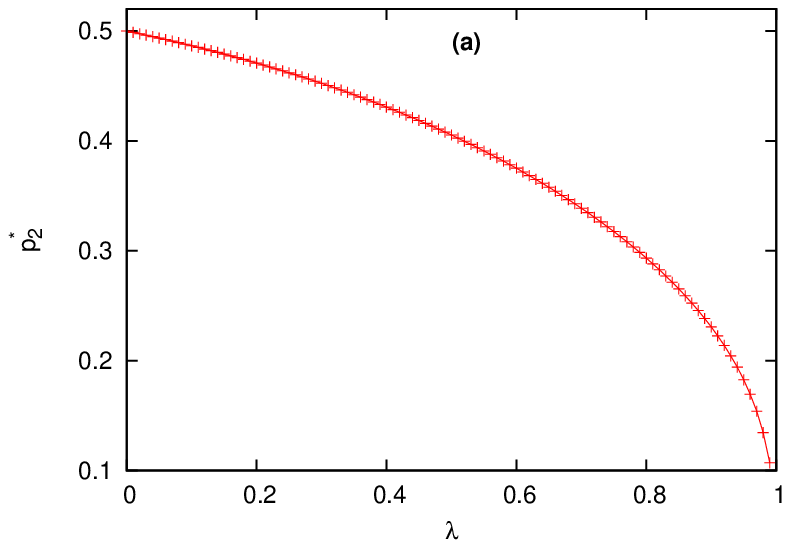}
 \includegraphics[scale = .68]{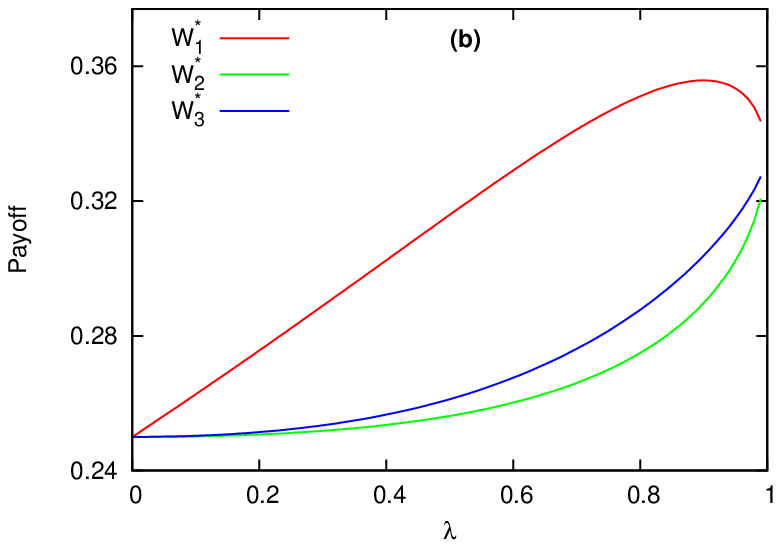}
\includegraphics[scale = .68]{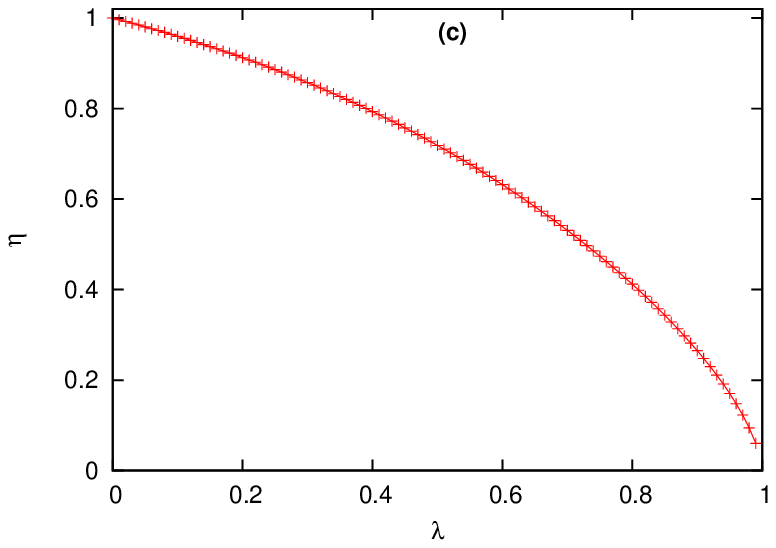}
 \caption{$N = 3$: (a) Variation of $p_2^*$ with $\lambda$, (b) The optimum payoffs  $W_i^*$, ($i= 1$ to $3$), as functions of $\lambda$ and (c) Inefficiency $\eta$ as a function of $\lambda$. .} 
\label{fig_1}
\end{figure*}

The variation of $p_2^*$ with $\lambda$ is shown in Fig \ref{fig_1}a.  $p_{2}^*$ monotonically decreases with $\lambda$ from its value $1/2$ at $\lambda = 0$, and tends to $0$ as $\lambda$ tends to $1$.  The pay-off of agents in various states with this optimum strategy is shown in Fig. \ref{fig_1}b and the variation of the inefficiency parameter $\eta$ with $\lambda$ is shown in Fig. \ref{fig_1}c.

It is easily seen that  $W_{avg}$ is a monotonically increasing function of $\lambda$, and tends  to the maximum possible value  $ W_{max} =1/3$ as $\lambda \rightarrow 1$. The variation of the inefficiency parameter $\eta$ with $\lambda$ is shown in Fig. \ref{fig_1}c. In particular, it is easily seen that $\eta$ varies as $(1 -\lambda)^{2/3}$, as $\lambda$ tends to $1$.

\subsection{{\bf \texorpdfstring{$N=5$}{}}}
We can similarly determine the optimal strategy for $N=5$. This is characterized by the five parameters $( p_1^*,p_2^*,p_3^*,p_4^*,p_5^*)$. The simplest strategy is ${\mathbb P}_{rand}$, which corresponds to  $p_i^* =1/2$, for all $i$. As explained above, the strategy ${\mathbb P}_{rand}' = ( 0,1/2,1/2,p_4^*(\lambda),1/2)$,  gives higher pay-off than ${\mathbb P}_{rand}$ for all agents, for all $\lambda$.

\begin{figure*}
\centering
 \includegraphics[scale = .65]{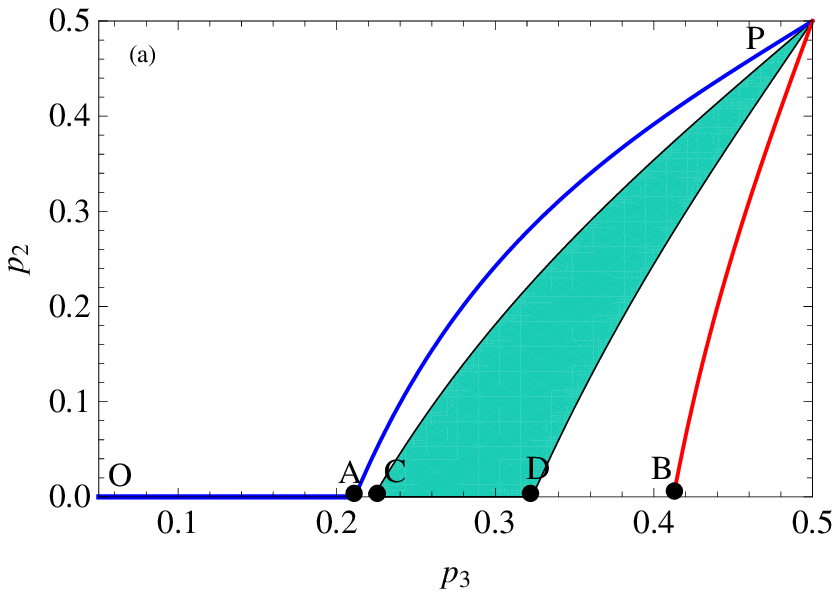} 
\includegraphics[scale = .65]{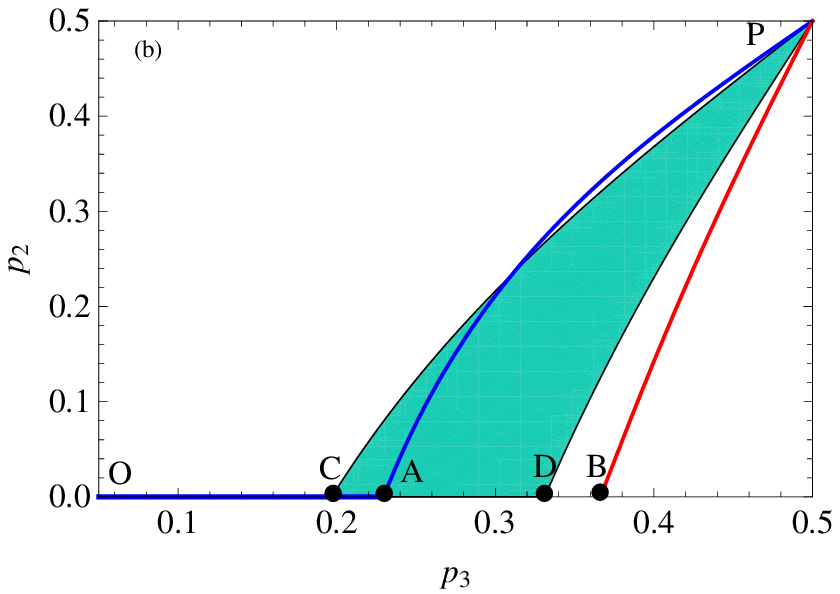}
\includegraphics[scale = .65]{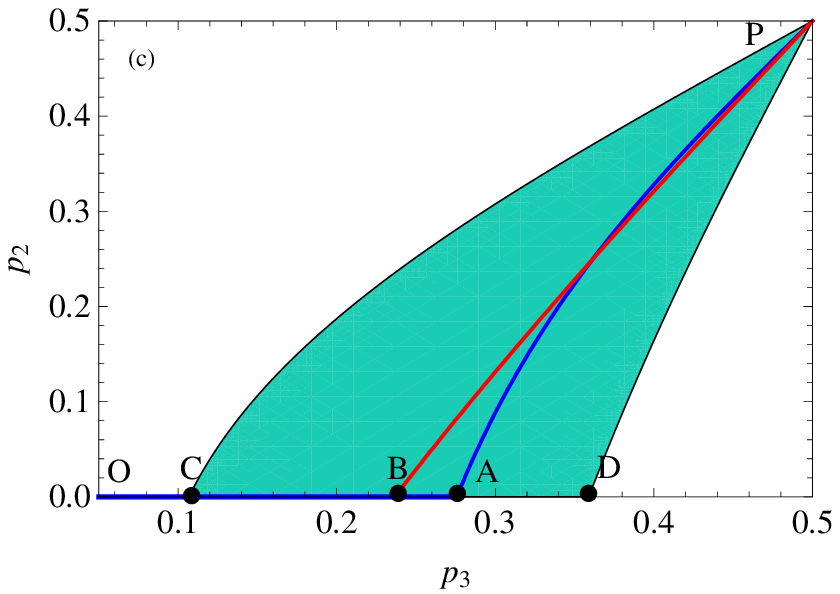}
\caption{Region in the $p_2$-$p_3$ plane showing the best responses $r_{2}^{opt}(p_3)$ (blue) and $r_{3}^{opt}(p_2)$ (red) for agents in state $|2\rangle$ and $|3\rangle$ respectively,  for (a) $\lambda =.1$, (b) $\lambda = .4$ and (c) $\lambda = .8$. The line $PC$ and $PD$ show the curves $w_3 = W'$  and $W_2 = W'$ respectively. In the curvilinear triangle $PCD$, all agents do at least as well as at $P$.}  
\label{P2P3}
\end{figure*}

Now consider agents in  the states $C_2$ and $C_3$. What  values of $p_2$ and $p_3$ they would select, given their expectation/belief about the selected values of $p_1$, $p_4$ and $p_5$ ?. We can determine these by  analyzing  the variation of payoffs $W_2$ and $W_3$ as  functions of $p_2$ and $p_3$ for fixed values of $p_1,p_4,p_5$ and $\lambda$ as  discussed below. 
  
Let us denote the best response of agents in state $C_2$ (that maximizes $W_2$),  if the agents in the opposite restaurant jump with probability $p_3$ by $r_2^{opt}(p_3)$. Similarly, $r_{3}^{opt}(p_2)$ denotes the best response of agents in state $C_3$, when those in the opposite restaurant jump with probability $p_2$. 

In Fig. \ref{P2P3}, we plot the functions $r_{2}^{opt}(p_3)$ ($OAP$) and $r_{3}^{opt}(p_2)$ ($BP$), in the $(p_3,p_2)$ plane, for three  representative values of $\lambda$. For small $p_3$, $r_{2}^{opt}(p_3)$ remains zero, and its graph sticks to x-axis initially, ( segment $OA$ in figure), and then increases monotonically with $p_3$.  The strategy ${\mathbb P}_{rand}'$ is the point $(1/2,1/2)$, denoted by $P$. We also show the lines $PC$ corresponding to $W_3 =W'$, and $PD$, corresponding to  $W_2 = W'$, where $W'$ is  the expected gain of agents in state $C_2$ or $C_3$ under ${\mathbb P}_{rand}'$. For all points in the curvilinear triangle $PCD$,  both  $W_2 $ and $W_3 \geq  W'$. Clearly,  possible equilibrium points are the points lying on the lines $r_{2}^{opt}(p_3)$, or $r_{3}^{opt}(p_2)$ that lie within the curvilinear triangle $PCD$. However,  along the blue curve $OAP$ representing $r_{2}^{opt}(p_3)$, maximum value for $W_2$ is achieved when $p_2 = 0$. Therefore we can restrict the discussion of 
possible equilibrium points to the line segment $CD$ in Fig. \ref{P2P3}.

For small $\lambda$ ( shown in  Fig. \ref{P2P3}a for $\lambda =0.1$), The point $A$ is to the left of $C$, and the only possible self-consistent equilibrium  point is $P$. 
For example, if agents in the state $C_3$ (Bob) assumes that agents in the minority restaurant (Alice) is going to set  $p_2^* =0$, Bob can get better pay-off than ${\mathbb P}'_{rand}$, by choosing his probability parameter $p_3^*$ in the range CD in Fig \ref{P2P3}a.  But  a rational Alice would not choose $p_2^* =0$, if she expects $p_3^*$ to be in the range CD. Similar argument rules out all points in the colored curvilinear triangle PCD as unstable.
This implies that  the agents would choose  $p_2^*=p_3^*=1/2$. This situation continues for all $\lambda < \lambda_{c1} = 0.195 \pm 0.001$.  

For $\lambda > \lambda_{c1}$, the point $A$ is to the right of $C$.  This is shown in Fig. \ref{P2P3}b, for $\lambda =0.4$. In this case, possible equilibrium points lie on the line-segment $CA$, and out of these, $A$ will be chosen by agents in state $C_3$. At $A$, both $W_2$ and $W_3$ are greater than $W'$, and hence this would be preferred by all. Further optimization  of $p_4$ changes $p_3$ and $p_4$  only slightly. 

\begin{figure*}
\centering
 \includegraphics[scale = 0.71]{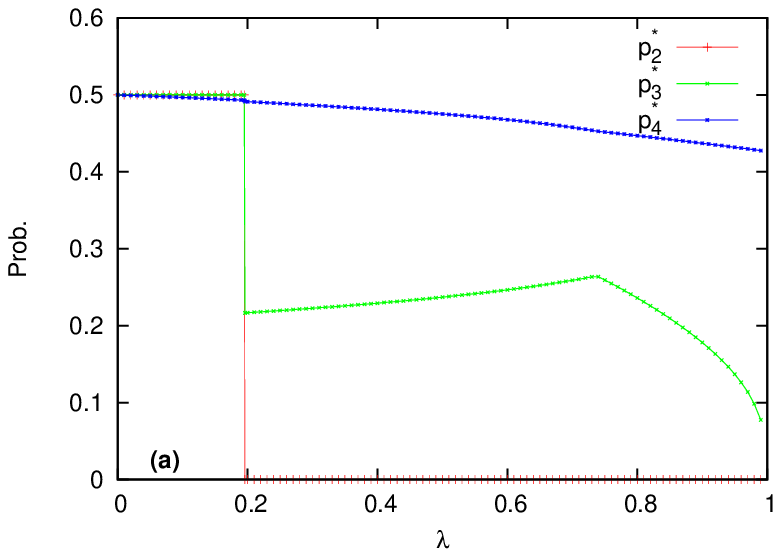} 
\includegraphics[scale = .71]{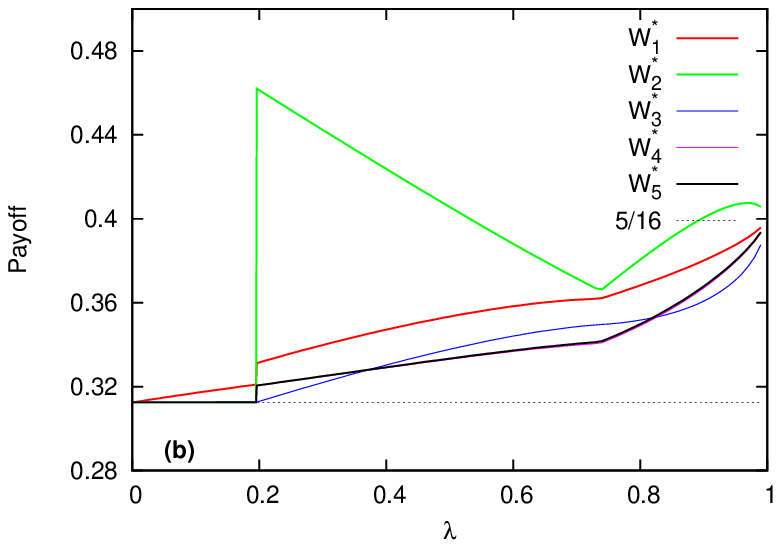}
\includegraphics[scale = .71]{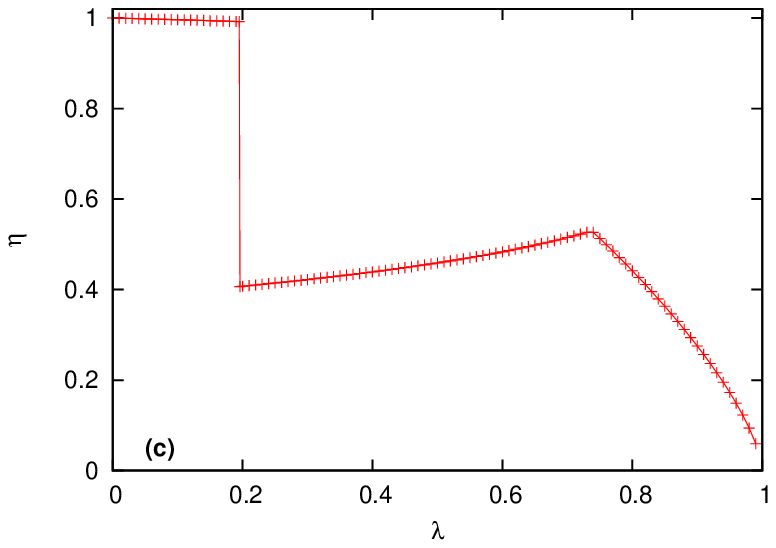}
\caption{$N=  5$: (a) Variation of $p_2^*$, $p_3^*$ and $p_4^*$ with $\lambda$ , (b) Optimum payoffs as functions of $\lambda$, (c) Inefficiency $\eta$ as a function of $\lambda$.}
\label{fig_3}
\end{figure*}

As we increase $\lambda $ further,  for $\lambda> \lambda_{c2}$ [numerically, $\lambda_{c2} = 0.737 \pm 0.001$], the point $B$ comes  to the left of $A$. Out of possible equilibria lying on the line-segment  $CA$, the point preferred by agents in state $C_3$ is no longer $A$, but   $B$. The self-consistent values  of $p_2^*$, $p_3^*$, and $p_4^*$ satisfying these conditions and the corresponding payoffs are shown in Fig. \ref{fig_3}a and Fig. \ref{fig_3}b  respectively.

In Fig. \ref{fig_3}c, we have plotted the inefficiency parameter $\eta$ as a function of $\lambda$.  For $\lambda < \lambda_{c1}$,  there are possible values of $p_2^*$ and $p_3^*$, that would increase the expected pay-off for everybody. However, Alice and Bob can not be sure that the other party would not take advantage of them, and hence stick to the  default sub-optimal-for-both choice $p_2^* = p_3^* = 1/2$.   

Also,  in the range $\lambda_{c1} < \lambda <  \lambda_{c2}$, the inefficiency rises as the agents optimize for farther into the future. This may appear paradoxical at first, as certainly, the agents could have used strategies corresponding to lower $\lambda$. This happens because though the game for  larger $\lambda$ is slightly less efficient overall, in it  the majority benefits more, as the difference between the optimum payoffs $W_2^*$ and $W_3^*$ is decreased substantially (Fig. \ref{fig_3}b).

We note  that the optimal strategies, and hence the (non-equilibrium) steady state of the system shows a non-analytic dependence on $\lambda$, even for finite $N$.  This is in contrast to the case of  systems in thermal equilibrium, where mathematically sharp phase transitions can occur only in the limit of infinite number of degrees of freedom $N$.  This may be understood by noting that the fully optimizing agents  in CAMG  make it more like   an  equilibrium system at zero-temperature. However note that unlike the latter, here the system shows a lot of fluctuations in the steady state. 

\subsection{{\bf Higher \texorpdfstring{$N$}{}}}

\begin{figure*}
\centering
 \includegraphics[scale = 0.71]{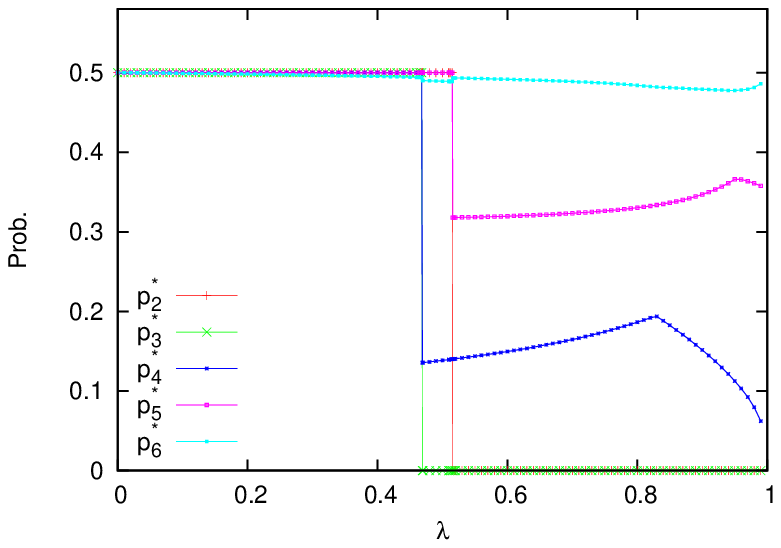} 
\includegraphics[scale = .71]{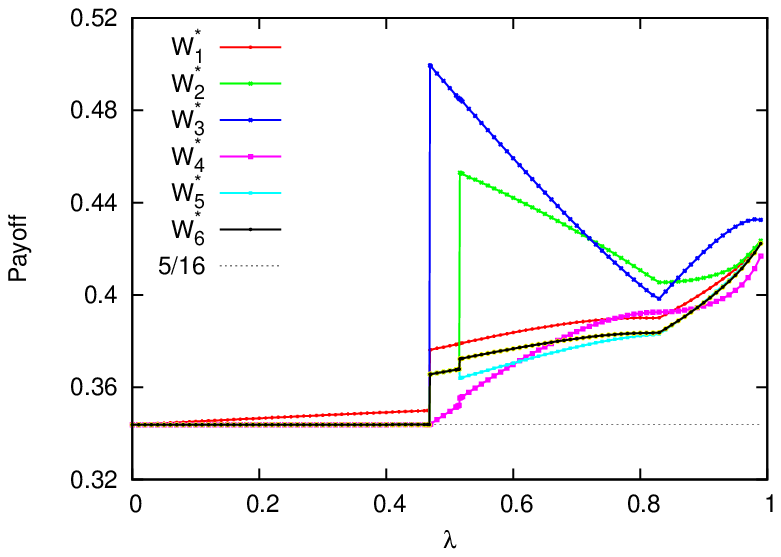}
\includegraphics[scale = .71]{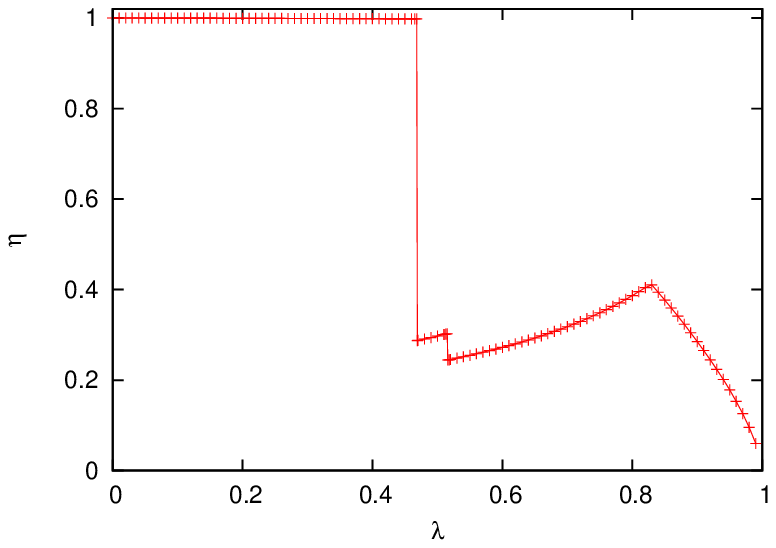}
\caption{$N=  7$: (a) Variation of optimum switch probabilities with $\lambda$ , (b) Optimum payoffs as functions of $\lambda$, (c) Inefficiency $\eta$ as a function of $\lambda$.}
\label{fig_4}
\end{figure*}

For higher values of $N$, the analysis is similar. For the case $N=7$, we find that there are four  thresholds $\lambda_{ci}$,
with $i= 1 $ to $4$. For $\lambda < \lambda_{c1}$, the optimal strategy has the form $(0,1/2,1/2,1/2,1/2,p_6^*,1/2)$. For $\lambda_{c1} \leq \lambda \le \lambda_{c2}$, we get $p_3^*=0$, and $p_4^* < 1/2$. For still higher values $\lambda_{c2} < \lambda \leq \lambda_{c3}$, agents in   the states $C_2$ and $C_5$ also find it better to switch to a win-stay lose-shift strategy, and we get $p_2^*=0$, $p_5^* < 1/2$. The transitions at $\lambda_{c3}$ and $\lambda_{c4}$ are similar to the second transition for $N=5$, in the $(p_4, p_3)$ and $(p_5,p_2)$ planes respectively. Numerically, we find $\lambda_{c1} \approx 0.47, \lambda_{c2} \approx 0.52, \lambda_{c3} \approx 0.83$ and $\lambda_{c4} \approx 0.95$. We present some graphs for the solution for $N=7$. Fig. \ref{fig_4}a shows variation of the optimum switch probabilities in various states and Fig. \ref{fig_4}b shows the variation of the optimum payoffs. Fig. \ref{fig_4}c shows the variation of inefficiency with $\lambda$. The general structure of the optimum 
strategy is thus clear. As $\lambda$ is increased, it changes from random switching to a complete win-stay lose-shift strategy in stages.  


An interesting consequence of the symmetry between the two restaurants is the following: If there is a solution $\{p_i^*\}$ of the self-consistent equations, another solution with all payoffs unchanged can be obtained by choosing for any $j$,  a solution $\{{p_i^*}'\}$, given by ${p_j^*}' = 1 - p_j^*$, and ${p_{N-j}^*}' = 1 - p_{N-j}$, and $p_i'=p_i$, for all $i \neq j$ and $i \neq N-j$. How agents choose between these symmetry related  $2^{M}$ equilibria can only be decided by local conventions. For example, if all agents follow the  \textquoteleft Win-stay lose-shift\textquoteright\;  convention, this would  select a unique equilibrium point.

\section {The Large-\texorpdfstring{$N$}{} limit}
\label{sec8}

In this section, we discuss the transition from the random strategy ${\mathbb P}_{rand}$, with all $p_j = 1/2$, to the strategy ${\mathbb P}_1$, in which  with $p_M^*= p_{M+1}^* =1/2$, and $p_j = 1/2$, for all other $j$. We will determine the  value of $\lambda_{c1}(N) $ where this switch  occurs.

The difference between the average payoffs  in the strategies ${\mathbb P}_{rand}$  and ${\mathbb P}_{rand}'$ is only of order $2^{-N}$, and may be ignored for large $N$. 

  In calculating the expected payoffs for strategy ${\mathbb P}_1$, it is convenient to group the states of the system into three groups: $|M\rangle, |M+1\rangle$, and the rest. These will de denoted by $|e_1\rangle, |e_2\rangle$ and $e_3\rangle$ respectively.

The transition matrix ${\mathbb T}$ may be taken as a $3 \times 3$ matrix. We consider the case when $p_{M+1}$ is ${\cal O}(M^{-5/4})$. Then ${\mathbb T}_{21}$ is ${\cal O}(M^{-1/4})$. It is convenient to  write ${\mathbb T}_{21} = a M^{-1/4}$, and use $a$ as  variational parameter, rather than $p_{M+1}$.  We also write  $b  =  ( 1 - \lambda)  M^{3/4} $.  We consider the case where $a$ and $b$ are finite, and ${\cal O}(1)$.  The transition probabilities    ${\mathbb T}_{12} = {\mathbb T}_{21} =  a M^{-1/4}$, and ${\mathbb T}_{31} = {\mathbb T}_{32} = a^2 M^{-1/2}/2$, to leading order in $M$. Also ${\mathbb T}_{13}={\mathbb T}_{23}$ is the probability that, when all agents are jumping at random, the marked agent will find himself in the state $|M\rangle$, (equivalently in state $|M+1\rangle$). For large $N$, this is well-approximated by the Gaussian approximation, and keeping only the leading term, we write   ${\cal W}_{13}={\cal W}_{23}= c M^{-1/2}$, where $c = 1/\sqrt{\pi}$.  

 Therefore we can write the transition matrix ${\mathbb T}$, keeping  terms only up to  ${\cal O}(M^{-1/2}$)  as,

\begin{equation}
{\mathbb T} = \left[ \begin{array}{ccc}
  1 - a M^{-1/4} -\dfrac{a^2 M^{-1/2}}{2}  & a M^{-1/4} & c M^{-1/2}\\
  a M^{-1/4} & 1 - a M^{-1/4} - \dfrac{a^2 M^{-1/2}}{2} & c M^{-1/2}\\
 \dfrac{a^2 M^{-1/2}}{2} & \dfrac{a^2 M^{-1/2}}{2} & 1 - 2 c M^{-1/2}
 \end{array} \right].
\end{equation}

Using the symmetry between the states $|e_1\rangle$ and $|e_2\rangle$, it is straight forward to diagonalize ${\cal W}$. Let the eigenvalues be $\mu_i$, with $i =1,2,3$, and the corresponding left and right eigenvectors be $\langle L_{i}|$ and $|R_{i}\rangle$.

For the steady state eigenvalue $\mu_1 =1$, we have  
\begin{equation} 
\nonumber \langle L_{1}| = \left[ 1,1,1\right];  ~~|R_{1}\rangle = \frac{1}{a^2 +4c} 
\left[\begin{array}{c}
2c\\
2c\\
a^2
\end{array}\right].
\end{equation}

The second eigenvalue is $\mu_2 = 1 - \frac{a^2 +4c}{2} M^{-1/2}$, and we have
\begin{equation} 
\nonumber \langle L_{2}| =   \frac{1}{a^2 +4c} \left[ a^2,a^2,-4c\right]; ~~~|R_{2}\rangle = \left[\begin{array}{c}
1/2\\
1/2\\
-1
\end{array}\right].
\end{equation}

The third eigenvalue is $\mu_3 = 1 - 2 a M^{-1/4} - a^2 M^{-1/2}/2$, and we have 
\begin{equation} 
\nonumber \langle L_{3}| =   \left[ 1/2, -1/2, 0\right]; ~~~|R_{3}\rangle = \left[\begin{array}{c}
1\\
-1\\
0
\end{array}\right].
\end{equation}

It is easily verified that $\langle L_{i}| R_{j}\rangle = \delta_{ij}$.

Now, we calculate the expected values of the payoff.  We note that if an agent is in the state $|e_3\rangle$, not only her 
exact state is uncertain, but even  her expected payoff depends on whether she reached this state from $|e_3\rangle$ in the previous day, or from $|e_2\rangle$.  This is because  the expected payoff in this state  depends on previous history of agent. However, her expected payoff {\it next day} depends only on  her current state (whether $|e_1\rangle$ or $|e_2\rangle$ or $|e_3\rangle$).

The expected payoff vector for the next day is easily seen to be 
\begin{equation}
\left[ W_{e_1}^{(0)},W_{e_2}^{(0)},W_{e_3}^{(0)}\right] = \left[ 1 - a M^{-1/4}  - a^2 M^{-1/2}/2,~ a M^{-1/4} + a^2 M^{-1/2}/2, ~1/2 - d M^{-1/2} \right],
\end{equation}
where $d = 1/(2\sqrt{\pi})$.
The expected payoff after $n$ days is given by $\left[ W_1^{(0)},W_2^{(0)},W_3^{(0)}\right] {\mathbb T}^{n-1}$. Then the discounted expected payoff with parameter $\lambda$ is given by
\begin{equation}
\left[ W_{e_1},W_{e_2},W_{e_3}\right] = \left[ W_{e_1}^{(0)},W_{e_2}^{(0)},W_{e_3}^{(0)}\right] \frac{(1 -\lambda)}{( 1 - \lambda{\mathbb T})}.
\end{equation}
 We write 
\begin{equation}
{\mathbb T} = \sum_{i=1}^{3} |R_i\rangle \mu_i \langle L_i|,
\end{equation}
and hence write 
\begin{equation}
\left[ W_{e_1},W_{e_2},W_{e_3}\right] = \sum_{i=1}^{3} U_i \langle L_i| ,
\end{equation}
where
\begin{equation}
U_i =  \left[ W_{e_1}^{(0)},W_{e_2}^{(0)},W_{e_3}^{(0)}\right] |R_i\rangle \frac{(1 - \lambda)}{( 1 - \lambda \mu_i)}.
\end{equation}

Now, explicitly evaluate $U_i$. We see that $U_1$ is independent of $\lambda$, and is the expected payoff in the steady state. The terms involving $M^{-1/4}$ cancel, and we get 
\begin{equation}
U_1 = \frac{1}{2} - \frac{d  a^2}{(a^2 + 4 c)} M^{-1/2}.
\end{equation}

For $U_2$, we note that $\left[ W_{e_1}^{(0)},W_{e_2}^{(0)},W_{e_3}^{(0)}\right] |R_2\rangle $ is of order $M^{-1/2}$, and $\frac{(1 - \lambda)}{( 1- \lambda \mu_2)}$ is of order $M^{-1/4}$, hence this term does not contribute to order $M^{-1/2}$.

The third term is $U_3$. Here the matrix element $\left[ W_{e_1}^{(0)},W_{e_2}^{(0)},W_{e_3}^{(0)}\right] |R_3\rangle $ is ${\cal O}(1)$, and $\frac{(1 - \lambda)}{( 1- \lambda \mu_3)}$ is of ${\cal O}(M^{-1/2})$, giving
\begin{equation}
U_3 = ( b /2 a)  M^{-1/2} + {\cal O}(M^{-3/4}).
\end{equation}

Putting these together, we get that $W_{e_2}$ is given by
\begin{equation}
W_{e_2} = 1/2 + M^{-1/2} \left[ - \frac{b}{ 4 a}  - d + \frac{ 4 d c}{ a^2 + 4 c} \right] +{\cal O}(M^{-3/4}).
\end{equation}

The agents in state $|e_2\rangle$ will choose the value $a = a^*$ to maximize this payoff $W_{e_2}$ with respect to $a$. Hence we have 
\begin{equation}
b = \frac{32 {a^*}^3 d c}{ ({a^*}^2 + 4 c )^2}.
\label{eqb}
\end{equation}
For any given $b$, we can solve this equation for $a^*$.  Then, at this point, the expeced payoff $W_{e_2}$ is
\begin{equation}
W_{e_2} = 1/2 - d M^{-1/2}\left[ 1  - \frac{4   c (  4 c -{a^*}^2)}{( {a^*}^2 + 4 c )^2} \right].
\end{equation}

This quantity is greater than  the expected payoff in the fully random state, so long as $ {a^*}^2 < 4c$, i.e.
\begin{equation}
b <  b_{max} =  2 {\pi }^{-3/4}.
\end{equation}
Thus, we see that if $\lambda > 1 - b_{max} M^{-3/4}$, there exists a nontrivial solution $a^*(b)$ satisfying Eq.
(\ref{eqb}), with $(a^*)^2 < 4 c$, and then the strategy in which agents in state $C_M$ stay, and $C_{M +1}$ shift with a small probability is beneficial to all.  Note that the future time horizon of agents only grows as a sub-linear power of $M$, while in the large $M$ limit, in CZMG, the time-scales grow  (at least) linearly with $M$. 

This large $M$ limit is somewhat subtle, as there are three implicit time scales in the problem: The  average time-interval between transitions between the states $|e_1\rangle$ and $ |e_2\rangle$ is of ${\cal O}(M^{1/4})$ days. Jumps into the state $|e_3\rangle$ occur at time-scales of ${\cal O}(M^{1/2})$ days. Once in the  state $|e_3\rangle$, the system tends to stay there for a time of ${\cal O}(M^{1/2})$ days, before a fluctuation again brings it to the state $|e_1\rangle$ or $|e_2\rangle$. The third time scale of ${\cal O}(M^{3/4})$  is the minimum scale of future horizon of agents required if the small per day benefit of a more efficient steady state 
of ${\cal O}(M^{-1/2})$ is to offset the cumulative disadvantage to the agents in state $|e_2\rangle$ of ${\cal O}(M^{1/4})$.

Note that the above analysis only determines the critical value of $\lambda$ above which the strategy ${\mathbb P}_1$ becomes preferred over ${\mathbb P}_{rand}$. This would be the actual critical value of $\lambda$ if the transition to the win-stay-lose-shift occurs in stages, as is suggested by the small $N$ examples we worked out explicitly. However, we cannot rule out the possibility that for  $N$ much larger than 7, the shift does not occur in stages, but in one shot, and  such  a strategy (similar to  the one described in \cite{dhar})  may be preferred over ${\mathbb P}_{rand}$ at much lower values of $\lambda$.

\section{Summary and concluding remarks}
\label{sec9} 
In this paper, we have analyzed a stochastic variant of the minority game,  where the $N$ agents are equal (no quenched randomness in strategies given to agents). This permits an exact solution in terms of $N$ self-consistently determined parameters. The solution shows multiple sharp transitions as a function of the discount parameter $\lambda$, even for finite $N$.  The main reason for the improved efficiency is that random number generators used by agents are much more effective in providing controlled differentiation between them than scoring methods for strategies. Also, the agents actually optimize the value of jump probability, and not use some   preassigned noise parameter.  In general, the performance using the CAMG  is found to be better than  in CZMG.  Also, there is some numerical evidence, and a qualitative argument that the relaxation time to reach the steady state  increases rather slowly, roughly  as $\log N$, compared to time of order $N$ days in CZMG \cite{dhar}.

Our treatment of the model here differs from that in \cite{dhar}: in that paper, the game was discussed only for $\lambda=0$ (corresponds to agents optimizing only next day's payoff), and in terms of Nash equilibrium. Within the Nash solution concept, it was not clear how to avoid the problem of trapping states,  and we had made an ad hoc assumption   that  whenever the system reaches a trapping state, a major resetting event occurs where all agents switch restaurant with some largish probability. In the co-action equilibrium concept proposed here, the decision to switch to $p=1/2$ or not is made rationally by the agents themselves depending upon their future time horizon. 


  Generalizations of the model where agents look back further than last day  are easy to define,   but even in the case $N=3$, this already becomes quite complicated, involving a simultaneous optimization over 9 parameters.    Introducing  inhomogeneity in the agents, say agents with different time horizons, is much more difficult.  In such a game, even if agents knew what fraction use what  discount parameter, knowing  only the record of attendances, would have to guess the fraction in their restaurant, and this makes the problem much harder to analyse.  The technique can be used to study other games with different pay-off functions, e.g. agents win only when their restaurant has attendance exactly  equal to some specified number $r$,  and these  appear to be interesting subjects for further study.


\acknowledgments
We thank D Challet, K. Damle, P. Grassberger, M. Marsili and  J. Stilck  for their comments on an earlier version of this paper.

\end{document}